\documentclass[pre,aps,showpacs,preprint]{revtex4}

\newcommand{\beq}{\begin{equation}}
\newcommand{\eeq}{\end{equation}}
\newcommand{\bqn}{\begin{eqnarray}}
\newcommand{\eqn}{\end{eqnarray}}
\newcommand{\bqns}{\begin{eqnarray*}}
\newcommand{\eqns}{\end{eqnarray*}}
\newcommand{\bary}{\begin{array}}
\newcommand{\eary}{\end{array}}

\usepackage{graphicx}%
\usepackage{amsmath}

\begin{document}
\title{Imaging an off-plane shear wave source with two-dimensional phononic-crystal lens}
\author{Chen-Yu Chiang and Pi-Gang Luan}
\address{Wave Engineering Laboratory, Department of Optics and
Photonics, National Central University, Jhungli 320, Taiwan}

\begin{abstract}%
A two dimensional flat phononic crystal (PC) lens for focusing
off-plane shear waves is proposed. The lens consists of a
triangular lattice hole-array, embedded in solid matrix.
Self-collimation effect is employed to guide the shear waves
propagating through the lens along specific directions.
Dirichlet-to-Neumann Maps (DtN) method is employed to calculate
the band structure of the PC, which can avoid the problems of bad
convergence and fake bands automatically in the void-solid PC
structure. When the lens is illuminated by the off-plane shear
waves emanating from a point source, a subwavelength image appears
in the far-field zone. The imaging characteristics are
investigated by calculating the displacement fields explicitly
using the multiple-scattering method, and the results are in good
agreement with the ray-trace predictions. Our results may provide
insights for designing new phononic devices.
\end{abstract}
 \maketitle
Phononic crystals (PC) are artificial periodic elastic media made
for controlling the propagation of acoustic/elastic waves. The
propagating modes of the elastic waves in PC are Bloch waves, and
their dispersion relations form the phononic band structures,
containing passbands and bandgaps. Various structures for
reflecting, trapping, and guiding acoustic/elastic waves relying
on the existence of PC bandgaps have been proposed and studied
\cite{Kush,Eco,Tor,Ping,ZQZ,Vas,Liu,Fang,Hsiao}. Besides the
bandgaps, the phononic passbands are also useful and have
interesting properties. When operating in the linear region well
below the top of the first band (i.e., the long-wavelength
region), a PC behaves just like a homogeneous medium, and
lens-like devices can be constructed from it
\cite{Airborne,ZhenYe}. Recently, the study of wave propagation
behaviors beyond the long-wavelength region has attracted a great
deal of interest. It has been shown that a slab of PC operating at
an appropriate frequency can focus the acoustic waves emanating
from a point source to form a subwavelength image, overcoming the
diffraction limit \cite{XdZhang,Qiu}.

To achieve the slab imaging ability, either negative refraction or
anomalous refraction is required \cite{Sukhovich,Bucay, Hennion,
JL}, which means the direction of the averaged Pointying vector
$\langle {\bf S}\rangle$ must have a component antiparallel to the
Bloch wavevector ${\bf k}$ or some effective Bloch wavevector
${\bf k}_{ef}={\bf k-k_0}$, i,e., ${\bf k}_{ef}\cdot\langle {\bf
S}\rangle<0$. Here ${\bf k_0}$ stands for certain symmetry point
in the first Brillouin zone or on the zone boundary, such as the
$M$ point for the square-lattice structure of PC. Recently, it has
been pointed out that the PC band structure can also provide other
wave guiding mechanisms such as self-collimation \cite{LSChen,
Perez, Espinosa, Shi} and canalization \cite{JL,He}. Using these
mechanisms, waves can be guided along certain specific directions
inside the slab and thus some desired devices such as hyperlens
can be realized \cite{CTChan}.

Up to now most of the studies on the negative refraction or
subwavelength imaging phenomena adopted the PC that sustain
pressure waves, and the discussions on similar phenomena in solid
elastic media are relatively few \cite{Hennion,Bucay, JL}. In this
letter, we employ the self-collimation effect of PC to design a
flat lens for focusing off-plane pure shear wave. The lens
consists of a triangular lattice hole-array embedded in a solid
matrix. When the lens is illuminated by the off-plane shear waves
emanating from a point source, a subwavelength image appears in
the far-field zone. The imaging characteristics for a finite lens
are investigated by calculating the displacement fields using the
{\it Multiple Scattering} (MS) method \cite{ZhenYe,LuanChang}. The
PC band structure is obtained with the {\it Dirichlet-to-Neumann
Maps} (DtN) method \cite{Lu1,Lu2}. Effective negative refractive
index of the PC can be extracted from the equal frequency contour
(EFC), and the image points can be predicted according to Snell's
law. The results are in good agreement with the calculations of
field distributions.

It is known that in calculating the band structure of the
fluid-solid or void-solid PC, the traditional plane wave expansion
(PWE) method encounters the problems of bad convergence and fake
flat bands \cite{Gof,Hou}. These problems can be resolved by
replacing the void part with a ``pseudosolid" or ``low impedance
material" \cite{Laude, Vasseur}, which is an imagined solid
material having a very low density and appropriately chosen wave
speed (usually very large). When these material parameters are
properly chosen, the flat band will appear only in the
uninterested high frequency range, and thus we can safely obtain
the desired band structure in the moderate frequency range.
Although this modified PWE method can indeed avoid the flat band
problem, there is no definite way of choosing the material
parameters of the pseudosolid, and usually some experiences are
needed. In order to reduce the uncertainty, we adopt instead the
DtN method to calculate the band structure of the PC. Since this
method can be applied to the void-solid case directly, the flat
band problem disappears automatically. Up to now DtN method has
only been used in calculating the band structures or guided modes
of two-dimensional photonic crystal (PhC) systems, however, there
is no essential difference mathematically between the
two-dimensional (2D) electromagnetic (EM) and acoustic (AC)
systems if the polarizations of the waves are properly chosen. For
example, the transverse-magnetic (TM) wave propagating in a 2D PhC
and the pure shear wave propagating in a 2D solid PC are governed
by the same kind of scalar wave equation. Thus an EM wave problem
can be easily translated to an AC wave problem and a one-to-one
correspondence can be established between these two systems
\cite{LuanYang}. This consideration implies that DtN method can be
used in the present case of pure shear wave. A detailed
description of DtN method will be given later.

Since in our chosen structure the off-plane shear wave cannot
penetrate into the holes, the band structure of the PC is
determined only by the geometrical factors (the hole size, lattice
constant, and lattice type) and the material parameters (the mass
density and the Lam\'{e} constant for shear wave) of the host
medium. In fact, if the lattice type, hole shape and the filling
fraction of the hole in a unit cell are all given, the band
structure of the PC becomes universal \cite{Feng1,Feng2} if the
dimensionless frequency $\omega a/2\pi c_t$ is used as the
frequency variable. Here $c_t$ represents the propagation speed of
the shear wave in the host medium and $a$ is the lattice constant.
If we replace the original host medium by another material, the
universal band structure predicts that identical wave phenomena
can happen in this new PC, usually in a different frequency or for
a different PC size. Note that although for non-dispersive host
media the band structure for a PC consisting of several materials
can also be expressed in a size-independent universal manner, the
material-independence is a unique property of the hole-array PC.
Moreover, in this case the (shear) wave speed is the only relevant
material parameter to determine the band structure.

To be more specific, we choose aluminum as the host medium
($\rho=2.692{\rm g}/{\rm cm}^3$, $c_t=3220{\rm m}/{\rm s}$). The
lattice type is triangular, the lattice constant is $a$, and the
radius of the holes is $r_0=0.4a$. An off-plane monochromatic
shear wave ${\bf u}({\bf r},t)=\hat{\bf z}u({\bf r})e^{-i\omega
t}$ propagating (in plane) in the host medium satisfies the
Helmholtz equation \beq \nabla^2 u({\bf r})+k^2_t u({\bf
r})=0.\eeq Here $u({\bf r})=u(x,y)$ represents the amplitude of
the displacement, $k_t=\omega/c_t$ is the wave number of the shear
wave, and the $\nabla^2=\partial^2/{\partial
x}^2+\partial^2/{\partial y}^2$ is the 2D Laplacian operator
because $u$ does not depend on $z$. Since the holes are all
circular, displacement field in the host medium around a hole
(say, the ith hole) can be expressed as a sum of cylindrical
waves: \beq u({\bf r})=\sum^{\infty}_{n=-\infty}C_n\left(J_n(k_tr)
-\frac{J'_n(k_tr_0)}{Y'_n(k_tr_0)}Y_n(k_t
r)\right)e^{in\phi}.\label{ueq}\eeq Here $r=|{\bf r}-{\bf r}_i|$
is the magnitude of the position vector relative to the center of
the ith hole, and $\phi=\phi_{{\bf r}-{\bf r}_i}$ is its azimuthal
angle. To derive this expression, the boundary condition that the
shear stress vanishes along the hole boundary has been applied.

Equation (\ref{ueq}) says that both $u$ and its first derivative
alone any direction are determined by the coefficients $C_n$.
Therefore, a DtN map between $u$ and its normal derivative
$\partial u/\partial n$ along the boundary of a unit cell can be
established. Using this relation together with Bloch boundary
conditions $u({\bf r+a_1})=e^{i{\bf k}\cdot{\bf a}_1}u({\bf r})$,
$u({\bf r+a_2})=e^{i{\bf k}\cdot{\bf a}_2}u({\bf r})$, the Bloch
wave modes $u({\bf r})$ and the PC band structure can be
determined \cite{Lu1,Lu2}. Here ${\bf a}_{1,2}$ represent the
primitive lattice vectors, and ${\bf k}$ is the Bloch wave vector.
Equation similar to (\ref{ueq}) but expanded as sum of Bessel and
Hankel functions can also be derived. Using that equation,
together with the fact that the incident wave around a hole comes
from the scattered waves of other holes and the waves from the
sources, a matrix equation relating the scattering coefficients
and the source coefficients can be established. Solving this
matrix equation then gives us the displacement field in the host
medium under the illumination of the sources
\cite{ZhenYe,LuanChang}.

The band structure and the EFC of the PC calculated using DtN
method are shown in Fig.1 (a) and 1(b). We have also calculated
the same band structure using the pseudosolid method, and no
visible difference between the results of these two methods can be
found. The accuracy of the band structure is thus confirmed. The
two inclined straight lines in Fig. 1(a) stand for the ``sound
lines" of the host medium. To focus the shear waves emanating from
a point source, we choose $k_ta/2\pi ={\omega}a/2\pi c_t=0.6$ as
the operating (reduced) frequency. The EFC of the PC at this
frequency looks like a hexagon, which occupies an area a little
smaller than the corresponding circular EFC of the host medium.
The frequency surface $\omega({\bf k})$ of the PC in the 2nd band
around the $\Gamma$ point has a negative curvature (negative
effective phonon mass), which leads to ${\bf k}\cdot{\bf v}_g<0$,
and thus negative refraction can occur. Here ${\bf
v}_g=\nabla_{\bf k}\omega$ is the group velocity, pointing in the
direction of the greatest rate of increase of $\omega$ and is
perpendicular to EFC. It can also be shown that the direction of
${\bf v}_g$ is the same as the direction of energy flow (averaged
Poynting vector) \cite{Sakoda}. When a plane wave of wave vector
${\bf k}$ is incident upon a PC, the refracted wave vector ${\bf
k'}$ can be determined by three rules. First, the incident and
refracted waves have the same frequency. Thus the tips of ${\bf
k}$ and ${\bf k'}$ are located on the host and PC EFCs,
respectively. Second, the wave phase on the two sides of the
host-PC interface must be the same, which means the tangential
component of ${\bf k}$ is equal to that of ${\bf k'}$. Third, the
normal component of ${\bf v}_g$ does not change sign in the
refraction process, which is a consequence of the energy
conservation law $\nabla\cdot\langle{\bf S}\rangle=0$ for
monchromatic wave. An illustration of this refraction process in
terms of the ${\bf k}$-space description and its corresponding
${\bf r}$-space configuration is shown in Fig.1(b). The thick
black arrows represent the incident and refracted wave vectors,
whereas the pink arrows indicate the ${\bf v}_g$ (energy flow)
directions. Hereafter we always assume ${\Gamma}{\rm K}$ be the
orientation of the host-PC interface. The EFC plot in Fig. 1(b)
also indicates that if the incidence angle is in the interval
$(-\pi/3,\pi/3)$, wave can penetrate into the PC and the
refraction angle of ray (energy flow) are almost fixed at
$\pm\pi/6$. This is caused by the self-collimation effect of the
hexagonal shaped EFC. On the other hand, if the incident angle is
beyond $(-\pi/3,\pi/3)$, no refracted wave can be excited, and the
incident wave will be totally reflected. These phenomena are
illustrated in Fig.2 (a)-(c). Here we use Gaussian beams as
incidence waves instead of plane waves. The Gaussian source used
in these simulations is formed by giving a Gaussian strength
profile along $20$ aligned point sources distributed evenly within
$4a$. The incidence angles in Fig. 2(a), (b) and (c) are $\pi/18$
($10^\circ$) and $2\pi/9$ ($40^\circ$), and $\pi/3$ ($60^\circ$),
respectively. The $\pi/6$ bending refraction beams in the first
two cases and the almost totally reflected beam in the third case
are clearly observed.

The self-collimation effects in square lattice PC or PhC have
already been discussed by some researchers \cite{JL, LSChen,
Perez, Espinosa, Shi}, however, to our knowledge, this effect in
triangular lattice structures has not yet been utilized as a means
to design an acoustic lens for focusing sound. Here we show its
applicability. Unlike most previously designed flat lens using
circular EFC, we use a hexagonal EFC to achieve this goal. Usually
a size-matched circular EFC having an effective refraction index
close to $-1$ is preferred, because it has several advantages such
as that for any incidence angle a refracted wave can always be
excited, and all rays can be brought to the same focal spot.
However, usually size-matched EFC can only be realized through
sophisticated methods, such as using more than one kinds of host
media \cite{Sukhovich}. Our aim in this paper is to design a
simple PC lens without relying on those sophisticated methods,
thus the hole-array structure is our choice. In order to reduce
the total reflection power, the EFC size of the PC must be large
enough and close to that of the host medium. This consideration
leads us to choose $\omega a/2\pi c_t=0.6$ as the working
frequency. If a slightly higher frequency EFC is chosen, the
reflection gets higher, and the image intensity gets dimmer. The
anisotropic refraction property of the PC looks like a
disadvantage, but we will show that a flat lens of this PC can
indeed work very well and it can brought the waves emitted by a
point source to a image spot of subwavelength wide.

Now we put a point source in front of the lens and study the
imaging effect. Simulation results for 9-layer and 13-layer
structures are shown in Fig. 3(a)-(c) and (d)-(f), respectively.
From (a) to (c) or (d) to (f) the source-lens distance is
increased from $a$ to $6a$. According to these simulation results,
the shear waves inside the PC indeed propagate along specific
directions (the $\pm\pi/6$) as expected. Besides, the thicker the
lens, the farther the image locates, consistent with the negative
refraction description. Furthermore, when the source moves towards
the left, the image also moves towards the left, keeping the
source-image distance almost unaltered.

The intensities $|u|^2$ of the displacement field in the focal
planes (located at the positions of the image peaks) for the
9-layer and 13-layer lenses are shown in Fig. 4(a) and (b),
respectively. In each subplot three source-lens distances
($d=0.8a$, $3.3a$, and $5.8a$) are examined, and the image widths
$w$ defined by their full-width-at-half-maxima (FWHM) are
recorded. The results reveal that the lens can really make a
subwavelength image of the point source at the far field zone,
although the image width (about $0.6\lambda$) is a little larger
than that employing evanescent wave coupling mechanism \cite
{He,LuanNano}.

The position of focal point can be predicted by considering the
refraction of acoustic rays. Although the PC in the chosen
frequency cannot be represented by an isotropic effective medium,
a formal usage of Snell's law (the consequence of the continuity
of the tangential component of wave vector across the host-PC
interface) can still help us to find the position of the focal
point approximately. The effective refractive index $n=-0.8305$ of
the PC at reduced frequency $k_ta/2\pi=0.6$ can be derived from
the relation $n=-c_t|{\bf k}|/\omega=-|{\bf k}|/k_t$, here $k_t$
is the radius of the host EFC, and $|{\bf k}|$ is the shortest
radius (the line segment ${\Gamma {\rm P}}$ in Fig.1(b)) of the PC
EFC. We sketch the ray-trace diagram in Fig. 5. The refraction
angle are fixed at $\pm\pi/6$, as mentioned above. According to
Snell's law, we have $\sin\theta=n\sin(-\pi/6)$, which gives
$\theta=0.4282$ or $24.53^\circ$ The distances $x_1$ and $x_2$ are
related to $d_1$ and $d_2$ through the relation
$x_1/d_1=x_2/d_2=\tan\theta/\tan (\pi/6)=0.7906$. We thus have a
constant source-image distant $d_1+d_2+D=2.265D$, consistent with
the results obtained in the wave field simulations. This analysis
seems to imply only one incident angle is allowable, which is not
the case for waves emitted from a point source. The ray trace
calculated in this way represents the route having an extremum
phase aggregation, that is, waves interfere constructively along
this specific route. If a route slightly deviated from this
specific one were used, the phase propagation direction (the
direction of ${\bf k}$) and beam propagation direction (the
direction of $\langle {\bf S}\rangle$) inside the PC would not be
parallel to each other. This leads to partially destructive
interference and reduce the strength of the ray. A simple glance
of Fig. 3 makes these statements more concrete. The predicted
positions for the two images (one inside and one outside of the
lens) based on the ray formula and those obtained by wave
simulations were compared in Fig. 6. A good agreement between
these two methods is found, as can be easily checked.

In conclusion, we have proposed a 2D phononic crystal flat lens
embedded in solid matrix to achieve far-field subwavelength
imaging of off-plane shear waves. The field patterns were computed
by using the multiple scattering method, whereas the band
structure of the phononic crystal were obtained by using the
Dirichlet-to-Neumann map method. Image positions were derived by
using ray-trace method, and the results are in good agreement with
those from direct wave simulations. Our study has the following
novelties. First, our PC lens focuses transverse shear wave, while
most of previously proposed PC lenses focus pressure wave. Second,
although we choose aluminum as the host medium, the PC band
structure is in fact universal when properly normalized, thus the
same kind of lens can be fabricated using other solid matrix
materials. Third, we demonstrated that the self-collimation effect
of triangular lattice PC can be employed to focus elastic waves,
thus circular shaped equal-frequency-contour is not necessary for
achieving negative refraction or subwavelength imaging. Our study
may provide some insights for designing new kinds of phononic
devices.

\section*{ACKNOWLEDGEMENT} The author gratefully acknowledge
financial support from National Science Council (Grant No. NSC
95-2221-E-008-114-MY3) of the Republic of China, Taiwan.


\begin{figure}
    \includegraphics[width=6in]{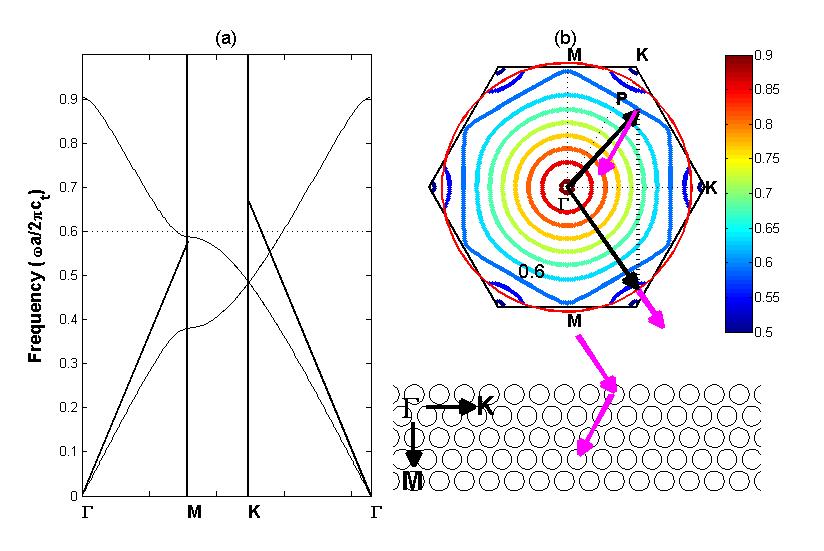}
    \caption{(a) The band structure of the triangular lattice hole-array PC. (b)
    The equal frequency contours (EFC) and the refraction diagram. The lattice constant is $a$,
    and the radius of the holes is $r_0=0.4a$.
    The two straight lines in (a) are the ``sound lines" for the shear wave. The
    hexagonal blue curve marked 0.6 and the red circle in (b) represent the EFCs of the PC
    and of the host medium respectively, at reduced (dimensionless) frequency
    $k_ta/2\pi=\omega a/2\pi c_t=0.6$. The two thick black arrows indicate the ${\bf k}$ vectors of the incident and
    refracted waves. The continuity of the wave phase along the host-PC interface implies that these two
    ${\bf k}$ vectors have the same component along the interface. The two pink arrows represent the energy flow (averaged Poynting vector) directions of the incident and
    refracted waves. The wave vector length $|{\bf k}|=\Gamma {\rm P}$ is chosen for calculating the effective
    refraction index $n$. Note that the direction of the refracted ray (energy flow) is approximately
    fixed at $\pm \pi/6$. Besides, it is not possible to excite a propagating mode of the PC if the incident
    angle is larger than $\pi/3$.}
\end{figure}

\begin{figure}
    \includegraphics[width=6in]{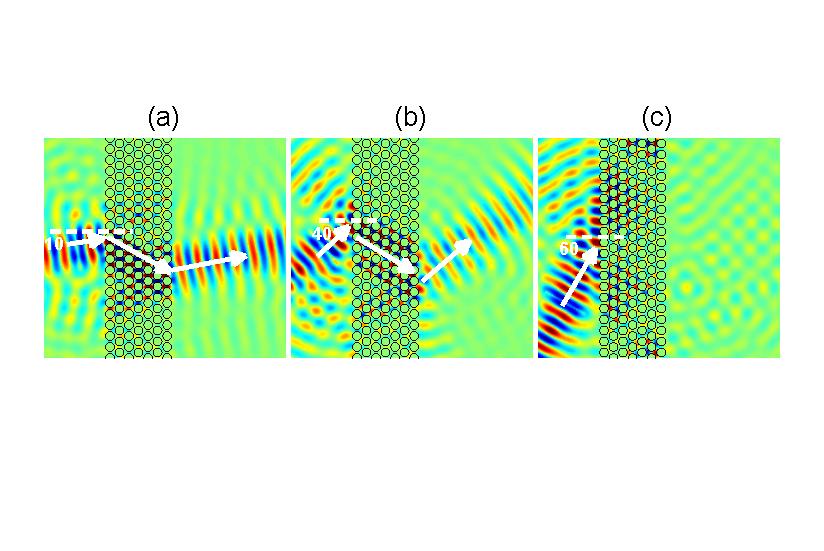}
    \caption{When a shear wave beam of (dimensionless) frequency $0.6$
    is incident upon the PC, it gets refracted with angle $\pi/6$
    (if the incidence angle is smaller then $\pi/3$) or totally
    reflected (if the incidence angle is close to or larger
    then $\pi/3$). The field patterns for incidence angles equal to
    $10^\circ$, $40^\circ$ and $60^\circ$ are shown in (a), (b) and (c), respectively.}
\end{figure}

\begin{figure}
    \includegraphics[width=6in]{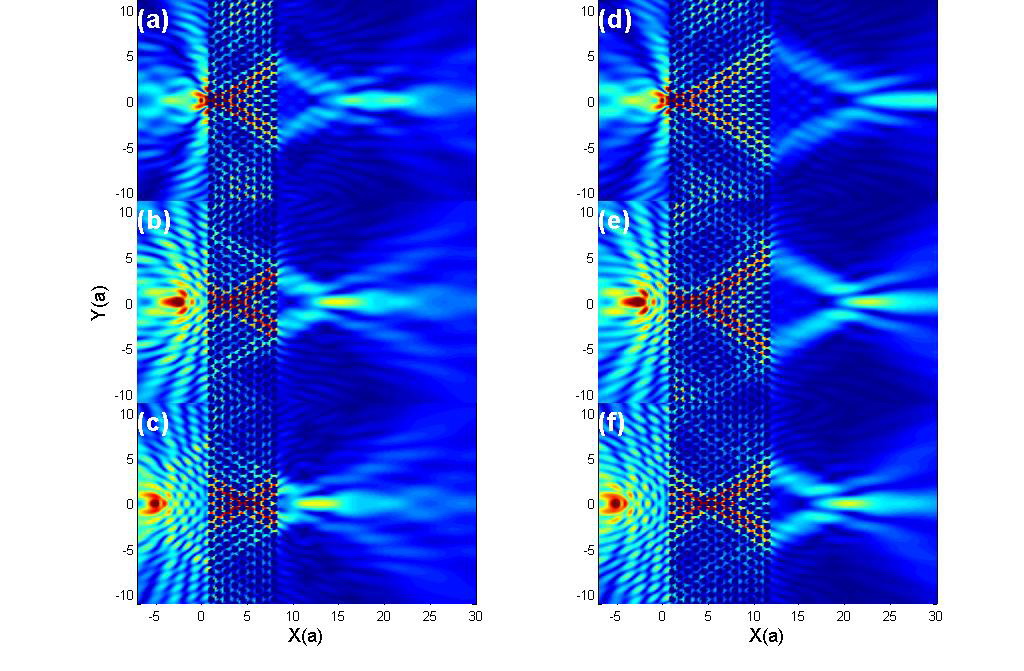}
    \caption{The displacement field patterns under the illumination of a 2D point
    source. The center of the first layer of holes is
    located at $x=a$. The results for the 9-layer structure are shown in (a)-(c).
    The source is located at $x=$ (a) 0, (b) $-2.5a$, and (c) $-5a$, respectively.
    Similar simulation results for the 13-layer structure are shown in (d)-(f).}
\end{figure}

\begin{figure}
    \includegraphics[width=4.5in]{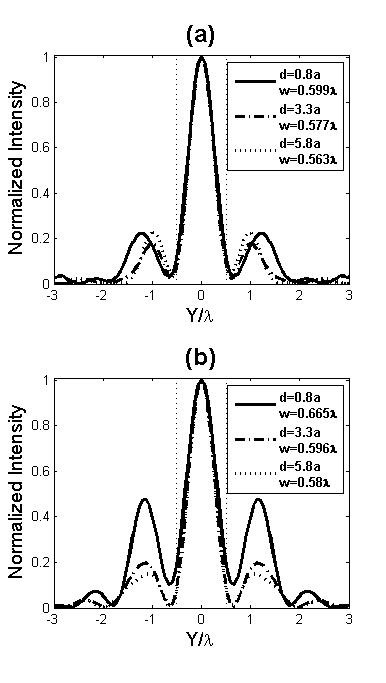}
    \caption{The shear wave intensities in the focal plane for the
    (a) 9-layer and (b) 13-layer structures. }
\end{figure}

\begin{figure}
    \includegraphics[width=6.5in]{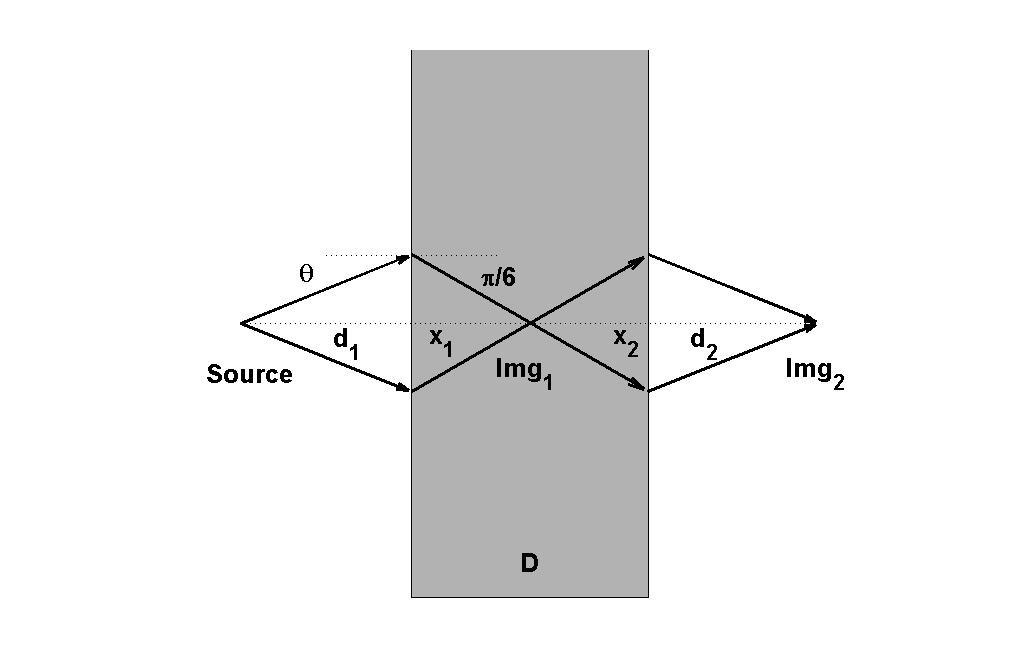}
    \caption{Refraction of acoustics rays. The refraction
    angles of $\pm \pi/6$ are expected inside the lens.}
\end{figure}

\begin{figure}
    \includegraphics[width=6.5in]{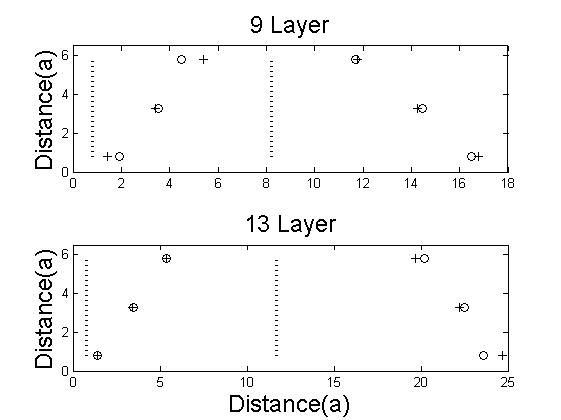}
    \caption{Comparisons between the predicted (the ``+" sign) and the actual (the ``o" sign)
    positions of the images. The vertical
    axis is the source-lens distance measured from the source location to the center of the first
    layer of the holes. The horizontal axis is the position of the
    image. The results for the 9-layer and 13-layer structures are
    shown in the upper and lower subplots, respectively.
    The two vertical lines represent the boundaries of the lens.}
\end{figure}

\end{document}